\documentclass{article}
\usepackage{amsmath}
\usepackage{amsfonts}
\usepackage{amssymb}
\usepackage{graphicx} 

\begin{document}


\title{How non-iridescent colors are generated by quasi-ordered structures of bird feathers}



\author{Heeso Noh$^{1}$, Seng Fatt Liew$^{1}$, Vinodkumar Saranathan$^{2}$, Richard O. Prum$^{2}$,\\ Simon G. J. Mochrie$^{1,4}$, Eric R. Dufresne$^{3,4}$ and Hui Cao$^{1,4}$\\ \\$^{1}$Department of Applied Physics\\$^{2}$Department of Ecology and Evolutionary Biology\\and Peabody National History Museum\\$^{3}$Departments of Mechanical Engineering,\\ Chemical Engineering and Cell Biology\\$^{4}$Department of Physics\\Center for Research on Interface Structure and Phenomena\\Yale University, New Haven 06511}

\date{}


\maketitle

\begin{abstract}
We investigated the mechanism of structural coloration by quasi-ordered nanostructures in bird feather barbs. Small-angle X-ray scattering (SAXS) data reveal the structures are isotropic and have short-range order on length scales comparable to optical wavelengths. We performed angle-resolved reflection and scattering spectrometry to fully characterize the colors under directional and omni-directional illumination of white light. Under directional lighting, the colors change with the angle between the directions of illumination and observation. The angular dispersion of the primary peaks in the scattering/reflection spectra can be well explained by constructive interference of light that is scattered only once in the quasi-ordered structures. Using the Fourier power spectra of structure from the SAXS data we calculated optical scattering spectra and explained why light scattering peak is the highest in the backscattering direction. Under omni-directional lighting, colors from the quasi-ordered structures are invariant with the viewing angle. The non-iridescent coloration results from isotropic nature of structures instead of strong backscattering. By comparing the data to those of an periodic structure, we demonstrated the difference in structural coloration between the quasi-ordered structures and ordered structures. 
\end{abstract}

\section{Introduction}

Organismal color is produced by chemical absorption of light by pigment molecules or by the physical interactions of light with nanostructures in the integument \cite{prum_anatomy_2006,kinoshita_physics_2008}.  The structural coloration is based on light scattering from nanoscale variation of refractive index in biological structures. Over the past decade there have been revived interests in structural coloration in nature, partly due to rapid developments of artificial photonic materials such as photonic crystals. The structural colors can be divided into two classes: (i) iridescent, (ii) non-iridescent. The iridescence means ``glittering or flashing with colors which change according to the position from which they are viewed''\cite{dic}. This definition assumes the objects are illuminated more or less uniformly from all directions, which is the usual case of lighting in nature. The iridescent color, e.g. from a butterfly wing, is generated by Bragg scattering of light from an ordered array of scatterers. Since the spatial distribution of scatterers is periodic, scattered light waves have correlated phase relationships, and the differential reinforcement or interference among them selects the color. The color changes with the angle of observation because the constructive interference condition is met at different wavelength for light scattered into different direction \cite{prum_anatomy_2006,ghiradella_structure_1989,vukusic_quantified_1999,kinoshita_structural_2005}. 

The non-iridescent colors from the feathers of many birds are produced by quasi-random arrays of air vacuoles in the medullary keratin. Because they lacked iridescence, these colors were hypothesized over 100 years ago to be produced by differential scattering of light wavelengths by individual scatterers \cite{Mason1923,Fox1976,finger_visible_1995}. Namely, the colors were dominated solely by the scattering properties of individual air vacuoles instead of constructive interference of light. For example, the blue color exhibited by feathers of numerous birds was attributed to the Tyndall effect. According to the Rayleigh law, the scattering efficiency of an air vacuole with size much smaller than wavelength is inversely proportional to the fourth power of wavelength. For larger scatterers, Mie scattering was considered to be the cause of structural color. 

In 1935 Raman threw doubt on the above hypothesis after conducting a thorough study of {\it Coracias Indica}, a common bird in Southern India whose feather exhibits non-iridescent color under natural light \cite{Raman1934}. Employing directional illumination from a light source, he observed striking variation of color from the bird wings with their positions relative to the source of light and the observer. Such variation could not be explained by Rayleigh or Mie scattering of individual air cavities embedded in keratin. He speculated the color was produced by constructive interference of light waves in extended structures. However, when the nanostructures were first seen in the 1940s following the invention of electron microscopes, their non-crystalline structures were interpreted as random. A lack of correlations in the spatial locations of scatterers implied the phase relationships among light waves scattered from them were random and could be ignored. Thus the single particle scattering hypothesis was upheld for non-iridescent color generation.  

Thirty years later, Raman's hypothesis was revived by Dyck who falsified the Rayleigh model for blue coloration. He documented that the reflectance spectra of many bird feathers showed discrete peaks in the visible spectra, which contradicted the prediction of continued intensity increase into UV by the Rayleigh model \cite{dyck_structure_1971,dyck_structural_1976}. Dyck hypothesized that the constructive interference of light scattered by the ordered matrix of air vacuoles and keratin produced the color, but he could not tell whether the structure was ordered enough to produce wavelength-specific color. 

Prum et al. provided the first convincing evidence of short-range order by Fourier analysis of electron micrographs of medullary keratin\cite{prum_coherent_1998,prum_two-dimensional_1999,prum_coherent_1999}. Two-dimensional Fourier transform of the nanostructures exhibits rings revealing the structures are isotropic and ordered on length scales comparable to the optical wavelengths   \cite{kinoshita_structural_2005,prum_coherent_1998,prum_two-dimensional_1999,prum_coherent_1999,prum_coherent_2003,prum_fourier_2003,prum_blue_2004,shawkey_mechanisms_2006,shawkey_nanostructure_2003,shawkey_anatomical_2005,shawkey_significance_2006,shawkey_electron_2009}. From the characteristic length scales of structures, Prum et al. predicted the wavelengths of optical reflection peak, which agreed reasonably well with the measured values. It proved the colors originated from the structural order. Since the structures are far from perfectly ordered ones, they are called quasi-ordered structures. 

The unidirectional reflectance spectrometry performed by Prum et al. could not tell whether the reflection peaks shifted with the observation angle. Osorio and Ham conducted a detailed study on the directional properties of structural coloration in bird plumage by measuring the reflectance spectra in various viewing geometries \cite{osorio_spectral_2002}. Under directional light illumination, the reflectance spectra depended both on the viewing angle and the feather orientation. The reflection peak wavelength varied linearly with the angle between the incident light and reflected light. Under diffusive light illumination, the reflectance spectra were insensitive to orientation and alignment of the feather. These results led to the question whether the colors from quasi-ordered structures are iridescent or non-iridescent.

Although the work of Prum et al. is decisive to the acceptance of light interference as the origin of color from the quasi-ordered biological structures, it remains unclear how the color is produced by constructive interference in these structures with only short-range order. More importantly, how can the color produced by optical interference be non-iridescent? Is the color truly non-iridescent, or under what conditions it is?  Answering the above questions is crucial not only to understand the physical mechanism of non-iridescent color generation by the quasi-ordered biological structures, but also to provide a clue why light interference had been mistaken for Rayleigh/Mie scattering for nearly a century.

In this paper, we investigated the physical mechanism of color generation by the quasi-ordered nanostructures of bird feathers. The small-angle X-ray scattering data confirmed the structures are isotropic and have only short-range order on the length scale comparable to optical wavelength. We performed angle-resolved reflection and scattering spectrometry to fully characterize color. With white light illumination of feather barbs, we measured the scattered light intensity as function of wavelength, sample orientation, incident light direction and viewing angle. Under directional illumination, the colors indeed changed with the incident angle and the viewing angle, more precisely, they depend only on the angle between the directions of illumination and observation. To simulate the natural lighting condition, the feather barbs are also illuminated by white light from all directions. The colors do not vary with the viewing angle or the orientation of feather barbs. These results demonstrate the subtleties of non-iridescent colors from the quasi-ordered structures, namely, the colors are non-iridescent under diffusive light illumination like in nature, but iridescent under artificial directional lighting often used in laboratory. Unlike the color of an ordered structure what is produced mostly by specular reflection, a quasi-ordered structure scatters light strongly into all directions, thus we used light scattering spectra instead of reflectance spectra for full characterization of colors from the quasi-ordered structures.

We interpreted the directional properties of coloration using the structure Fourier spectra obtained from the X-ray scattering experiment. Our calculation of light scattering spectra reproduced the spectral shifts of scattering peaks with angles. We also explained why the backscattering peak is the strongest. With omni-directional lighting the scattering peak is broader than that with directional lighting and its center wavelength is shorter than that of backscattering peak. Thus the color under natural light is not determined by the backscattering, even though it is the strongest. We provided an explanation based on the solid angle available for illumination and the non-iridescence of color in the natural lighting condition attributed to isotropic nature of quasi-ordered structures.


\section{Morphology of Quasi-Ordered Nanostructures in Feather Barbs}

\begin{figure}[htbp]
\center
\begin{tabular}{c c}
\includegraphics[width=2in]{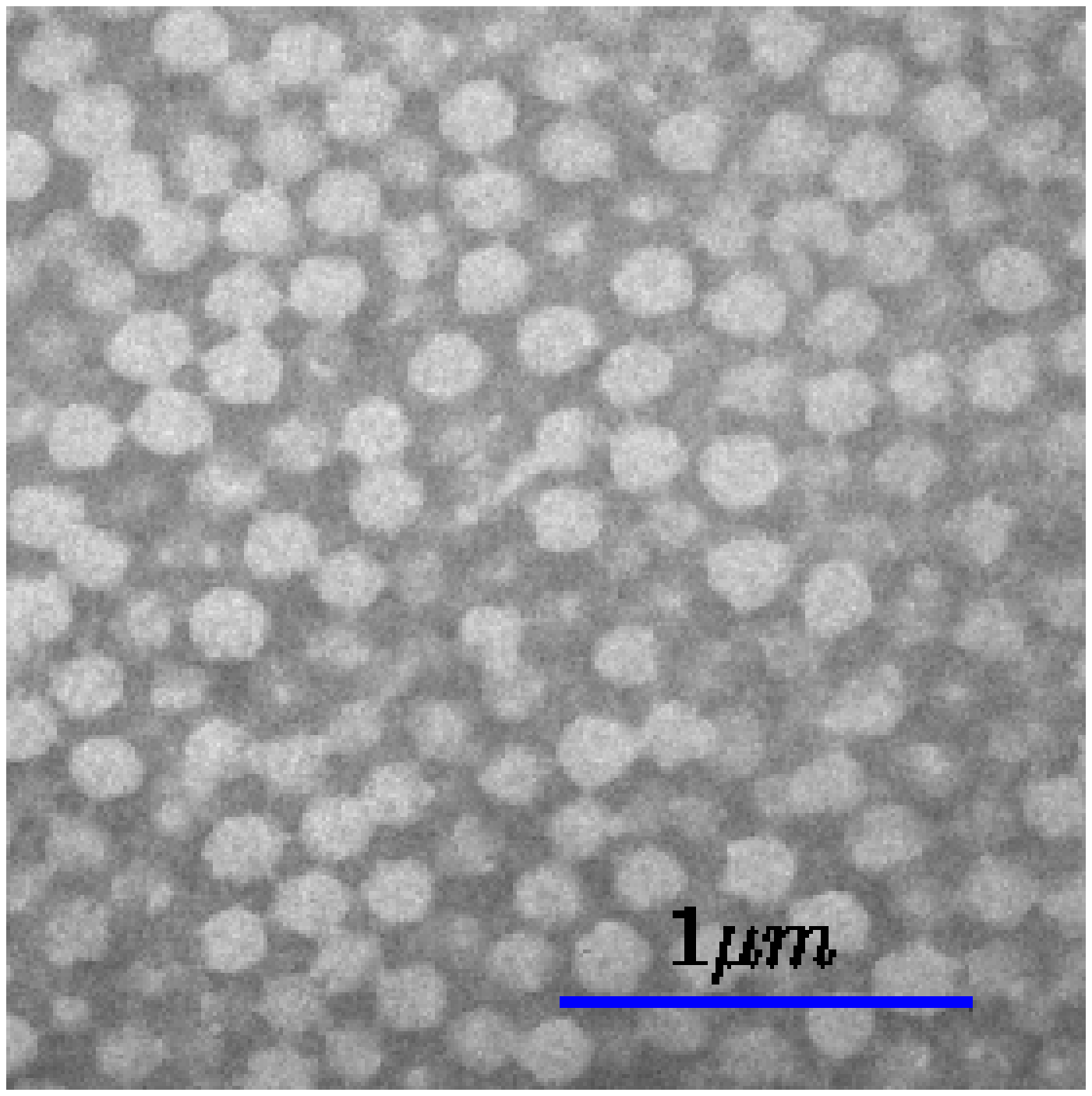}&
\includegraphics[width=2in]{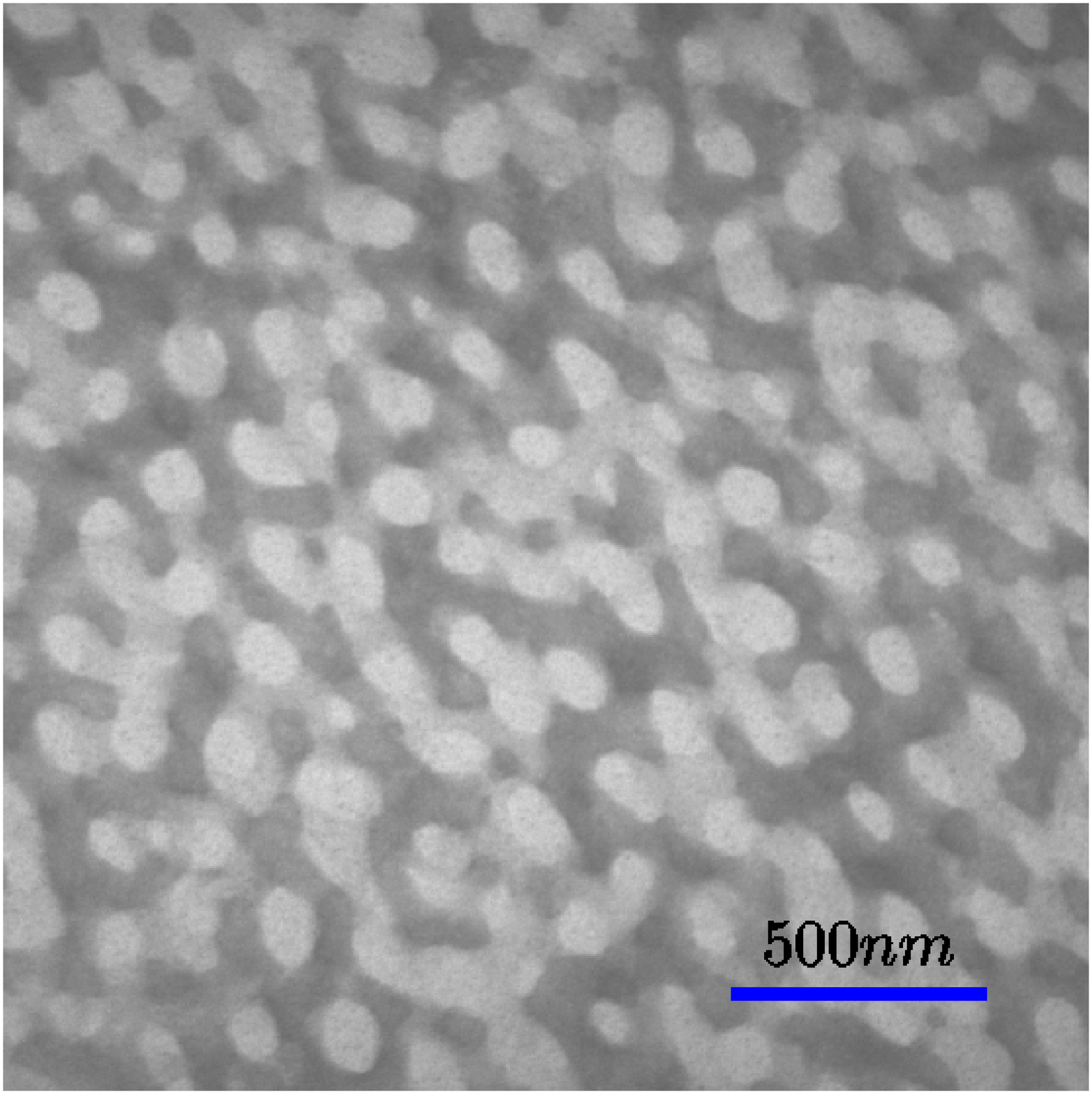}\\
(a)&(b)
\end{tabular}
\caption[TEM images of bird feather structures]{Transmission electron microscope (TEM) images of (a) sphere-type quasi-ordered nanostructures from {\it C. cotinga}, (b) channel-type $\beta$-keratin and air nanostructures from back contour feather barbs of {\it I. puella}. The dark-staining material is $\beta$-keratin and the unstained voids are air.}
\label{fig:tem}
\end{figure}

Structural colors are produced by the spongy $\beta$-keratin and air nanostructures within the medullary cells of bird feather barb rami \cite{dyck_structure_1971,dyck_structure_1971-1}. Known barb nanostructures belong to two distinct morphological classes \cite{prum_anatomy_2006,dyck_structural_1976}. Channel-type nanostructures consist of $\beta$-keratin bars and air channels in elongate, tortuous and twisting forms. Sphere-type nanostructures consist of spherical air cavities in a $\beta$-keratin background, sometimes interconnected by tiny air passages. TEM images of the two types of nanostructures are presented in Fig. \ref{fig:tem} (a) and (b). Our previous studies suggest these nanostructures are self-assembled during phase separation of $\beta$-keratin protein from the cytoplasm of the cell \cite{dufresne_self-assembly_2009,prum_development_2009}. The channel morphology is developed via spinodal decomposition and the sphere morphology, via nucleation and growth.  

We studied structural color production in the feather barbs of six avian species, three with channel-type structures, and three with sphere-type. In the following, we present the data for one channel-type, {\it Irena puella} ({\it I. puella}), and one sphere-type, {\it Cotinga cotinga} ({\it C.  cotinga}). 

\begin{figure}[htbp]
\center
\begin{tabular}{c c}
\includegraphics[width=3in]{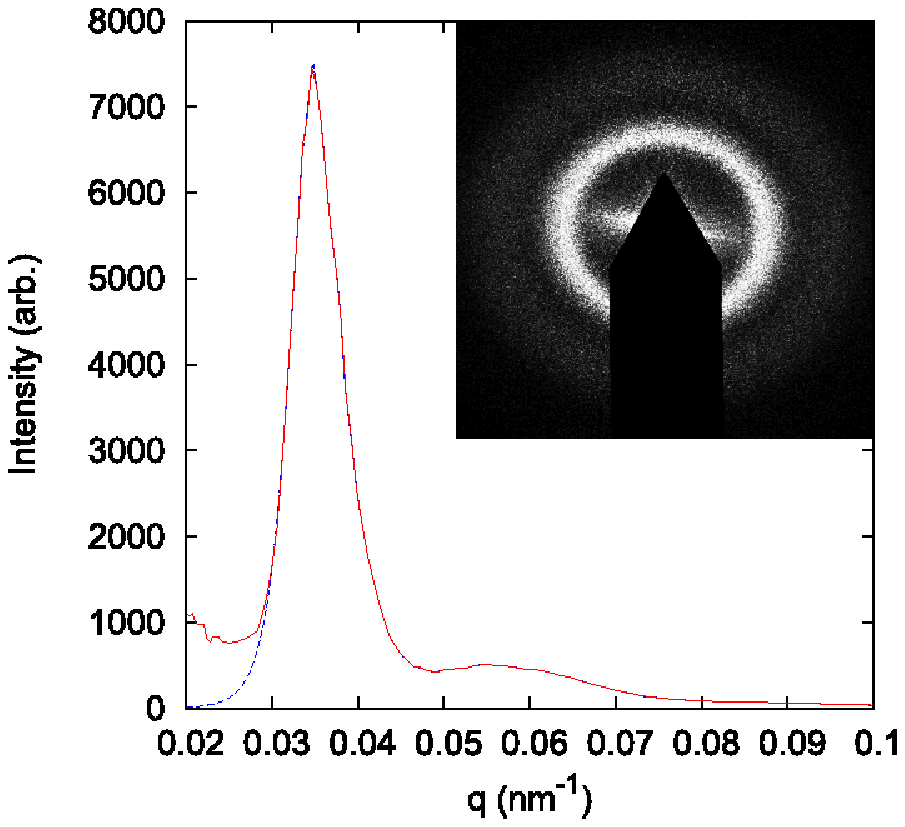}&
\includegraphics[width=3in]{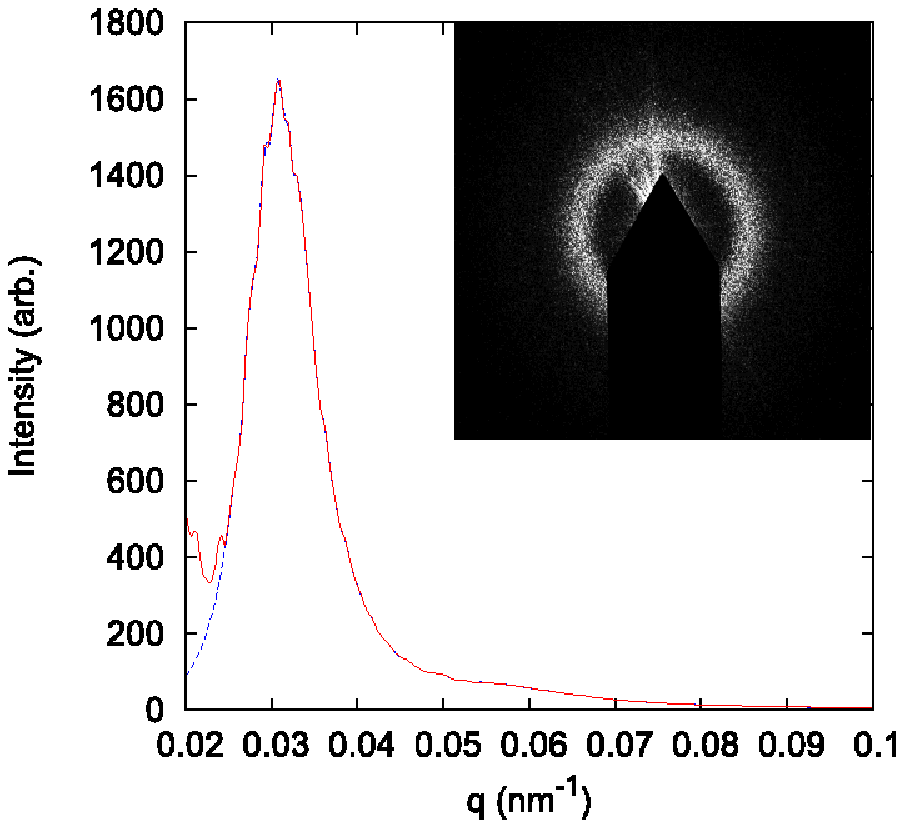}\\
(a)&(b)\\
\end{tabular}
\caption[angle averaged SAXS data]{Inset: Small-angle X-ray scattering (SAXS)
patterns from channel-type feather barb of {\it I. puella} (a) and sphere-type of {\it C. cotinga} (b). Main panel: azimuthal averages of SAXS data (red solid lines), and fitted curves (blue dashed lines). The curve fitting at smaller $q$ eliminates the artificial signal from diffraction of the beam block. }
\label{fig:SAXS}
\end{figure}

In addition to transmission electron microscopy, we measured small-angle X-ray scattering (SAXS) to characterize the nanostructures. SAXS directly measures the structural correlations via single scattering, without complications from multiple scattering. The X-ray scattering patterns for both the channel-type and sphere-type nanostructures exhibit rings [insets of Figs. \ref{fig:SAXS}(a) and (b)], implying that the structures are isotropic.  Exploiting the rotational symmetry of the SAXS pattern, we averaged the scattered X-ray intensities over the entire azimuthal angle to obtain the intensity $I$ as a function of spatial frequency $q$. As shown in Fig. \ref{fig:SAXS}, $I(q)$ has a primary peak, revealing the existence of a dominant length scale for structural correlations. The peak position $q_0$ gives the spatial correlation length $s= 2 \pi / q_0$. The full width at half maximum (FWHM) $\Delta q$ of the peak reflects the range of spatial order, $\xi = 2 \pi / \Delta q$. For {\it I. puella} $s$ = 204nm, and $\xi$ = 694nm. For {\it C. cotinga} $s$ = 182nm, and $\xi$ =906nm. The values of $s$ illustrate the nanostructures are ordered on length scales comparable to optical wavelengths. However, the order is short-ranged, because $\xi \sim 3s -5s$. The $I(q)$ for {\it C. cotinga} has a weaker second peak at higher spatial frequency, corresponding to a smaller correlation length. The second peak of $I(q)$ is not obvious for {\it I. puella}, and the main peak has a larger tail at higher $q$.   
We also obtained the spatial correlation functions of nanostructures from the Fourier transform of $I(q)$. They exhibit exponential decay with spatial separation, confirming the structures have only short-range order. 

\section{Optical Setup for Full Color Characterization}

\begin{figure}[htbp]
\center
\includegraphics[width=6in]{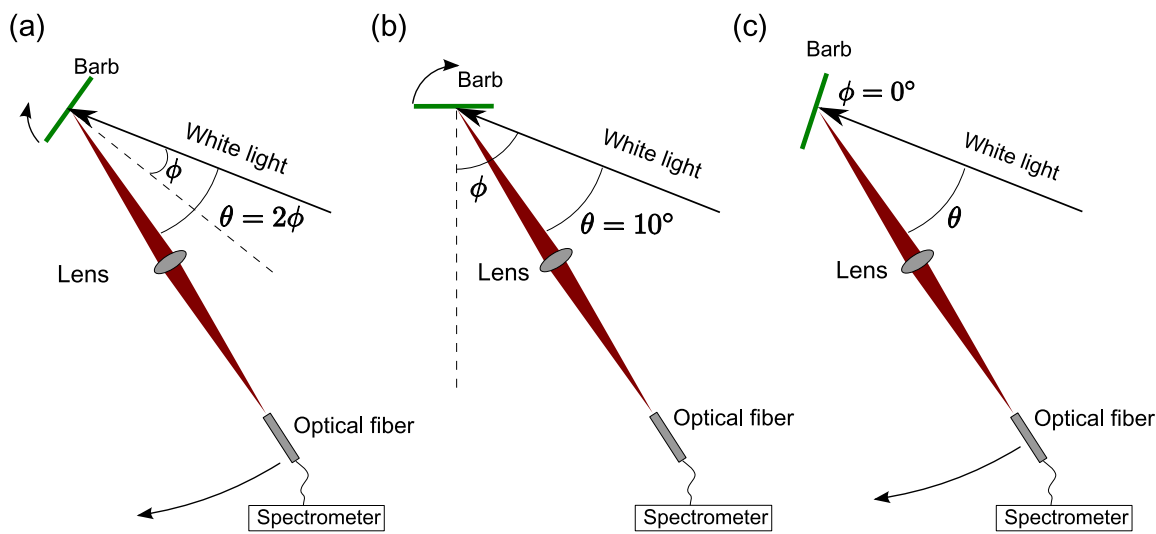}
\caption[experiment setup]{Schematic diagrams of optical setups for measurements of specular reflection spectra (a), near-backward scattering spectra (b), and diffusive scattering spectra (c) of white light.}
\label{fig:setup}
\end{figure}

We conducted angle-resolved reflection and scattering spectrometry to fully characterize the structural color. A bird feather is mounted at the rotation center of a  goniometer. Collimated white light from a UV enhanced Xe lamp is incident on the sample at an angle $\phi$ from the surface normal. The spot size on the sample is about 1 mm. Reflected or scattered light is  collected by a lens and focused to an optical fiber that is connected to a spectrometer (Ocean Optics HR2000+). The spectral resolution is 1.5 nm. The angular resolution, which is determined mainly by the collection angle of the lens, is about $5^{\circ}$. 

Both the sample and the detection arm on which the lens and fiber are mounted can be rotated. The specular reflection is measured in the $\phi - 2 \phi$ geometry shown in Fig. \ref{fig:setup} (a). Namely, when the sample is rotated angle $\phi$, the detection arm is rotated $2 \phi$. To measure the scattered light, we fix the sample and rotate only the detection arm [Fig. \ref{fig:setup} (c)]. In this case, the illumination angle remains constant, while the observation angle changes. We also fix the detection arm and rotate only the sample [Fig. \ref{fig:setup}(b)]. The angle $\theta$ between the incident beam and detection arm remains constant, while the angle of incidence $\phi$ changes with sample rotation. 

\begin{figure}[htbp]
\center
\includegraphics[width=2in]{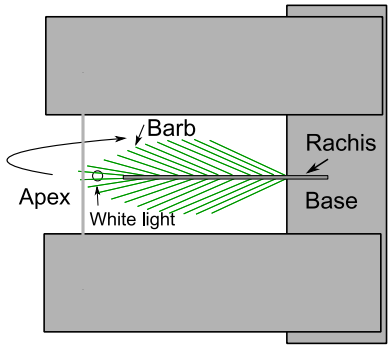}
\caption[sample mount]{ Schematic diagram showing the orientation of feather barbs with respect to the rotation axis of goniometer, the location of incident light beam and how the feather barbs are mounted with the probed area free of substrate.}
\label{fig:mount}
\end{figure}

The feather barbs with cylindrical cross-sections are mounted horizontally so that their axes are perpendicular to the rotation axis of the goniometer as shown in Fig. \ref{fig:mount}. If the barbs are mounted vertically, non-iridescent color might be produced via a different mechanism. Namely, if the periodic structures are arranged in concentric circular arrays conformal to the curved surfaces of barbs, the color would not change as the barbs rotate along their axes due to rotational symmetry. To tell whether the color is iridescent, we make the barbs rotate along an axis (i.e. the rotation axis of goniometer) perpendicular to their cylinder axial directions. In previous measurements \cite{osorio_spectral_2002}, the substrates on which the feathers were mounted were painted matte black to avoid reflection or scattering from the substrates. In our experiment, the apex and base of feather barbs are mounted on two separate holders while the probed parts in between do not have any substrate underneath. Thus the effects of substrates are completely removed from the reflection or scattering data. The measured spectra of reflected or scattered light are normalized by the spectrum of incident light. The normalized spectra are then smoothed by a convolution with a Gaussian function of full width at half maximum equal to 15 nm. 

\begin{figure}[htbp]
\center
\includegraphics[width=4in]{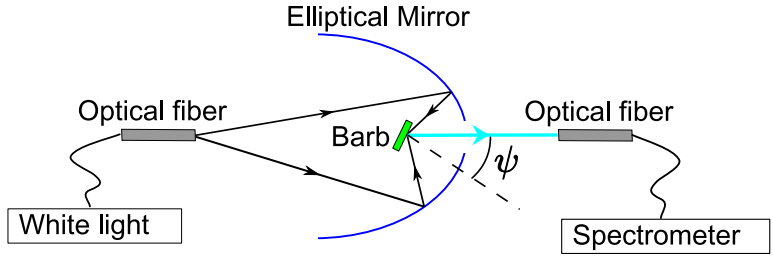}
\caption{Schematic diagram showing the optical setup for illuminating feather barbs with white light from all directions.}
\label{fig:totalsetup}
\end{figure}

Under natural lighting condition, the bird feathers are illuminated by ambient light coming in all directions. The structural color under natural lighting condition is likely to be different from that under directional lighting \cite{osorio_spectral_2002}. 
To simulate natural lighting, we use a setup shown in Fig. \ref{fig:totalsetup}. White light is coupled to one end of an optical fiber. The other end of fiber is placed at one focal point of an elliptical mirror. Light exiting the fiber end is focused by the mirror to the other focal point, where the feather barbs are located. The distance between the two focal points is 194 mm. A circular opening on the mirror allows light being scattered by the sample to a particular direction to be collected by another fiber that is connected to a spectrometer. The measured spectrum of scattered light is normalized to that of incident white light. The feather barbs can be rotated with respect to the detector. The angle between the direction of observation and the normal of barb surface is $\psi$. This setup simulates the situation of a bird watcher under natural lighting. 

\section{Iridescent Color under Directional lighting}

\begin{figure}[htbp]
\center
\begin{tabular}{c}
\includegraphics[width=3in]{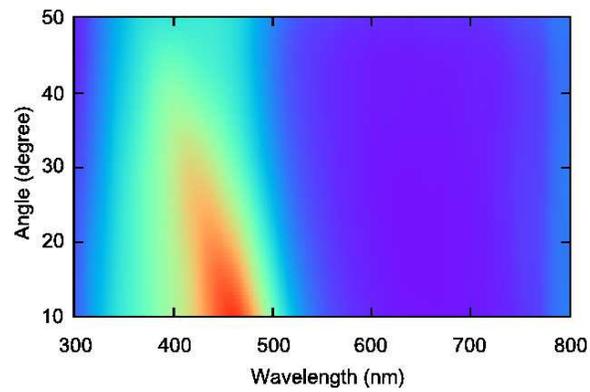}\\
(a)\\
\includegraphics[width=3in]{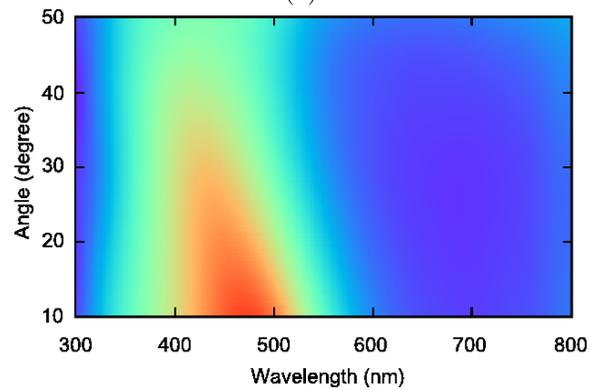}\\
(b)
\end{tabular}
\caption[Specular reflection]{False color maps showing the intensity of specularly reflected light as function of the incident angle $\phi$ and wavelength $\lambda$ from the feather barbs of {\it C. cotinga} (a), and {\it I. puella} (b). The experimental setup is shown in Fig. \ref{fig:setup} (a).}
\label{fig:specular}
\end{figure}

We started with the specular reflection measurement using the setup shown in Fig. \ref{fig:setup}(a). Figure \ref{fig:specular} depicts the false color maps showing the reflected intensity as function of the incident angle $\phi$ and wavelength $\lambda$ for three samples. Red corresponds to higher intensity, and blue to lower intensity. 
The quasi-ordered structures of both channel-type and sphere-type exhibit similar trend. The reflection peaks blue-shift with increasing $\phi$. These results indicate the colors generated by quasi-ordered structures depend on angles of incidence. This behavior is similar to that of an ordered structure.

\begin{figure}[htbp]
\center
\begin{tabular}{c}
\includegraphics[width=3in]{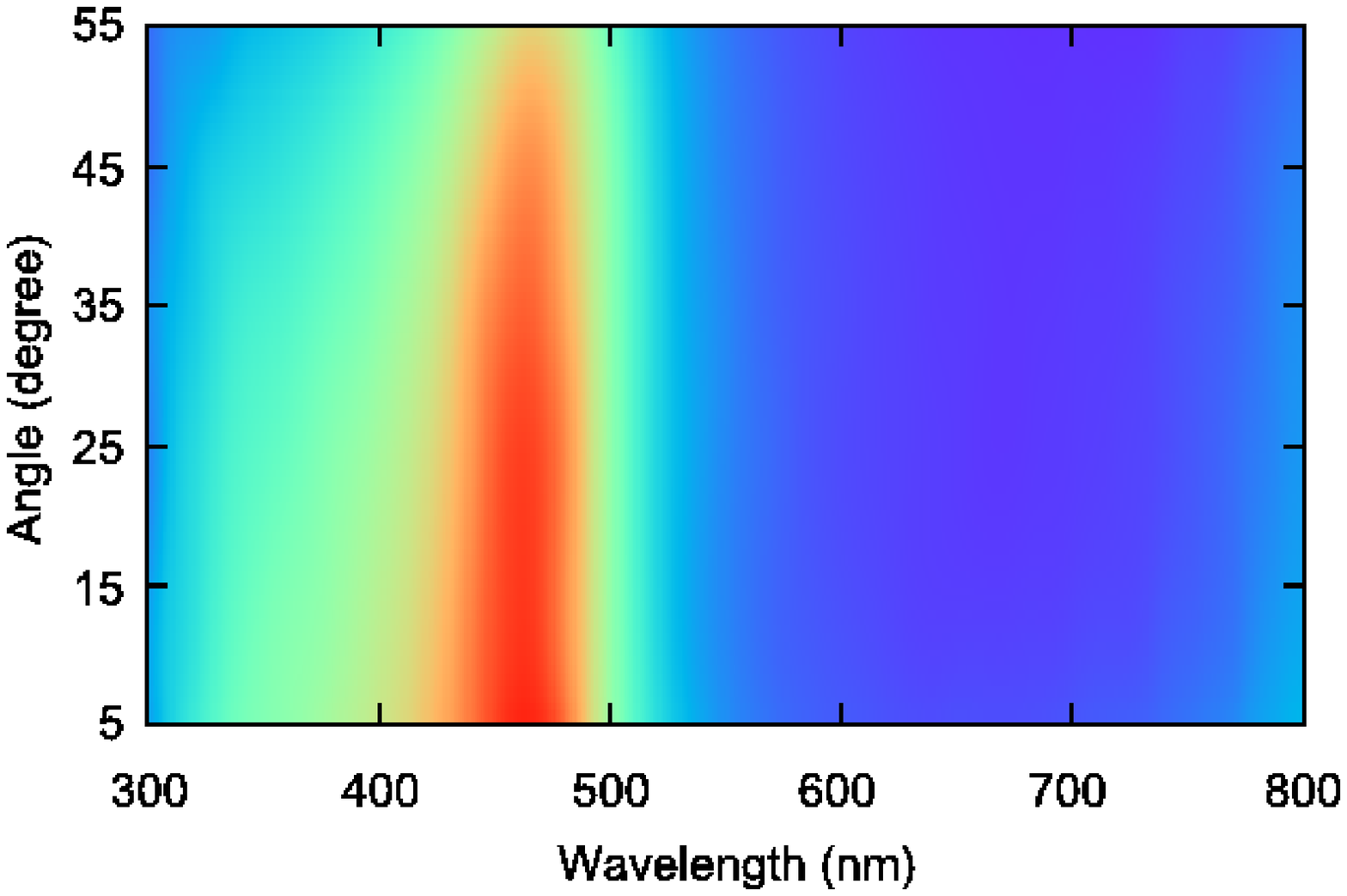}\\
(a)\\
\includegraphics[width=3in]{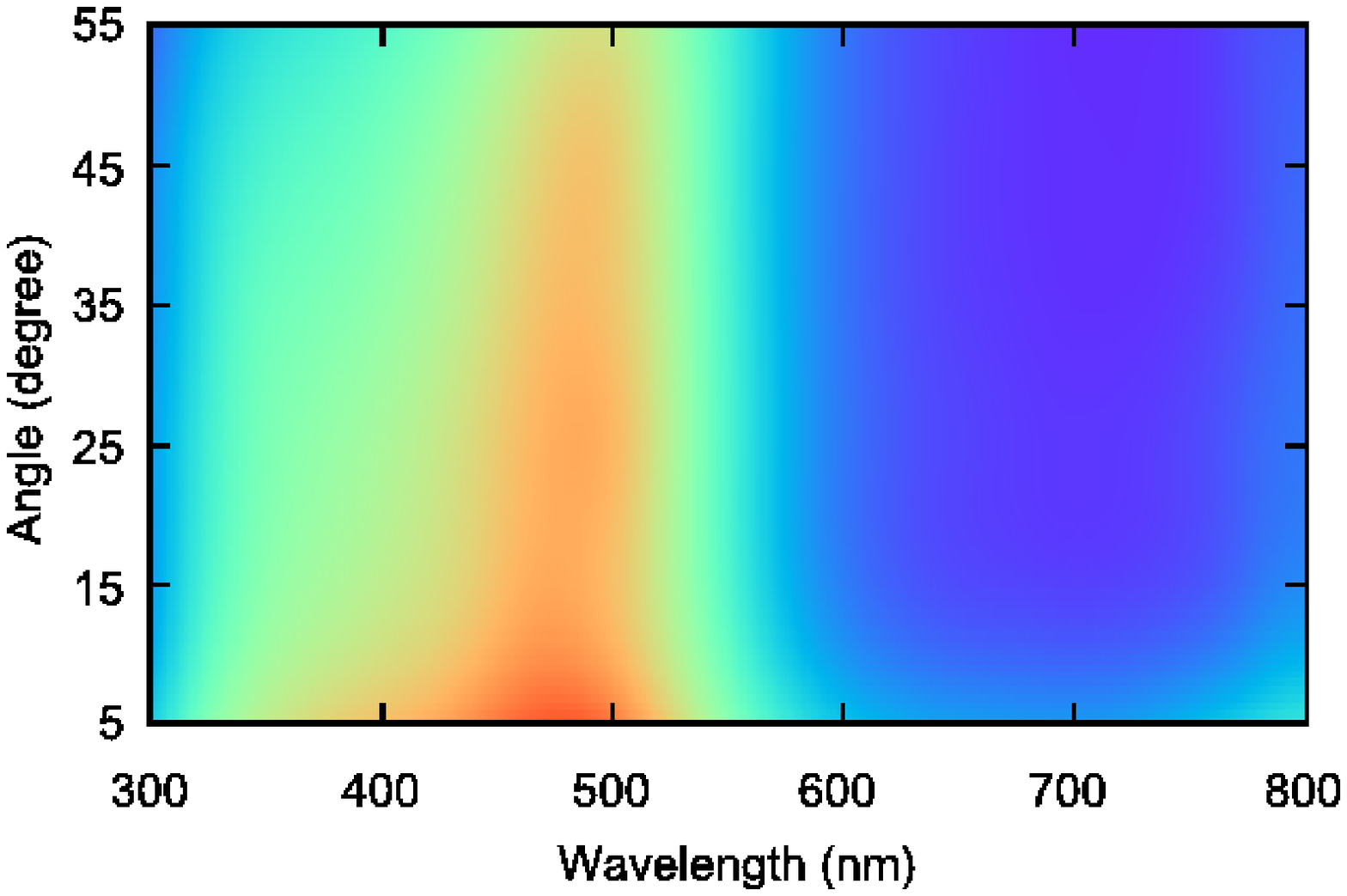}\\
(b)
\end{tabular}
\caption[Backward scattering]{False color maps showing the intensity of light scattered to nearly backward direction of the incident beam as function of the incident angle $\phi$ and wavelength $\lambda$ from the feather barbs of {\it C. cotinga} (a), and {\it I. puella} (b). The experimental setup is shown in Fig. \ref{fig:setup} (b). The angle $\theta$ between the directions of illumination and observation is fixed at $10^{\circ}$. }
\label{fig:backward}
\end{figure}

Next we fix the detection arm while rotating the sample [Fig. \ref{fig:setup}(b)]. The angle between the incident beam and detection arm $\theta$ is kept at 10$^{\circ}$. The incident angle $\phi$ varies from 5$^{\circ}$ to 55$^{\circ}$. Figure \ref{fig:backward} shows the measured intensity as function of $\phi$ and $\lambda$. 
For $\phi$ = 5$^{\circ}$, $\theta = 2 \phi$, thus the specularly reflected light enters the detector. The spectrum of {\it C. cotinga} has a primary peak at 462 nm, and that of {\it I. puella} at 478 nm.   As $\phi$ increases, the specularly reflected light moves away from the detector. At larger $\phi$ the detector collects the scattered light. As shown in Fig. \ref{fig:backward}, the spectral peaks remain for both types of quasi-ordered structures. This means light is not only specularly reflected, but also scattered into other directions. The peak wavelengths remain nearly constant as sample rotates. Thus the colors are independent of sample orientations, which is consistent with the isotropic nature of the quasi-ordered structures. 

\begin{figure}[htbp]
\center
\begin{tabular}{c}
\includegraphics[width=3in]{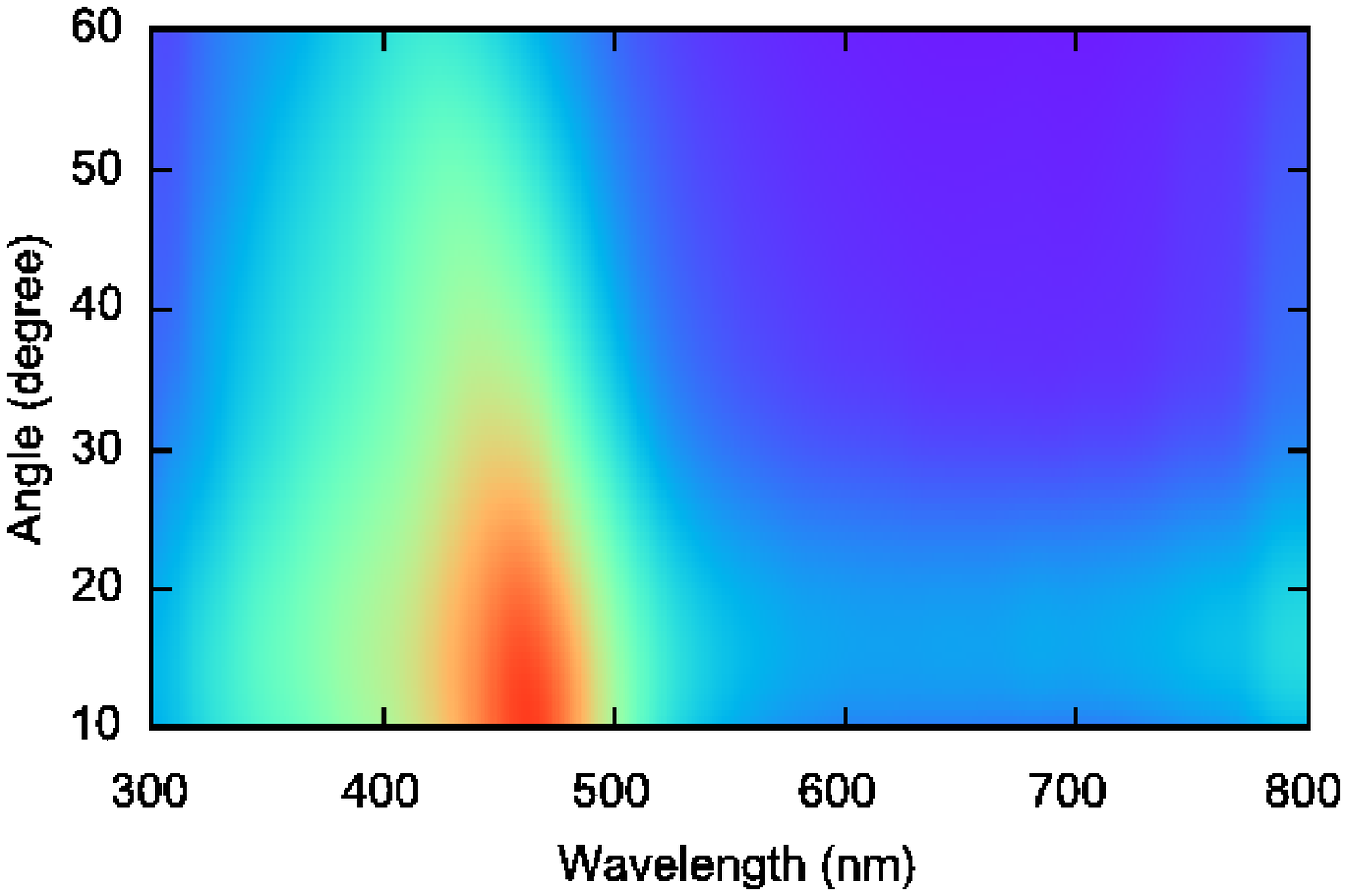}\\
(a)\\
\includegraphics[width=3in]{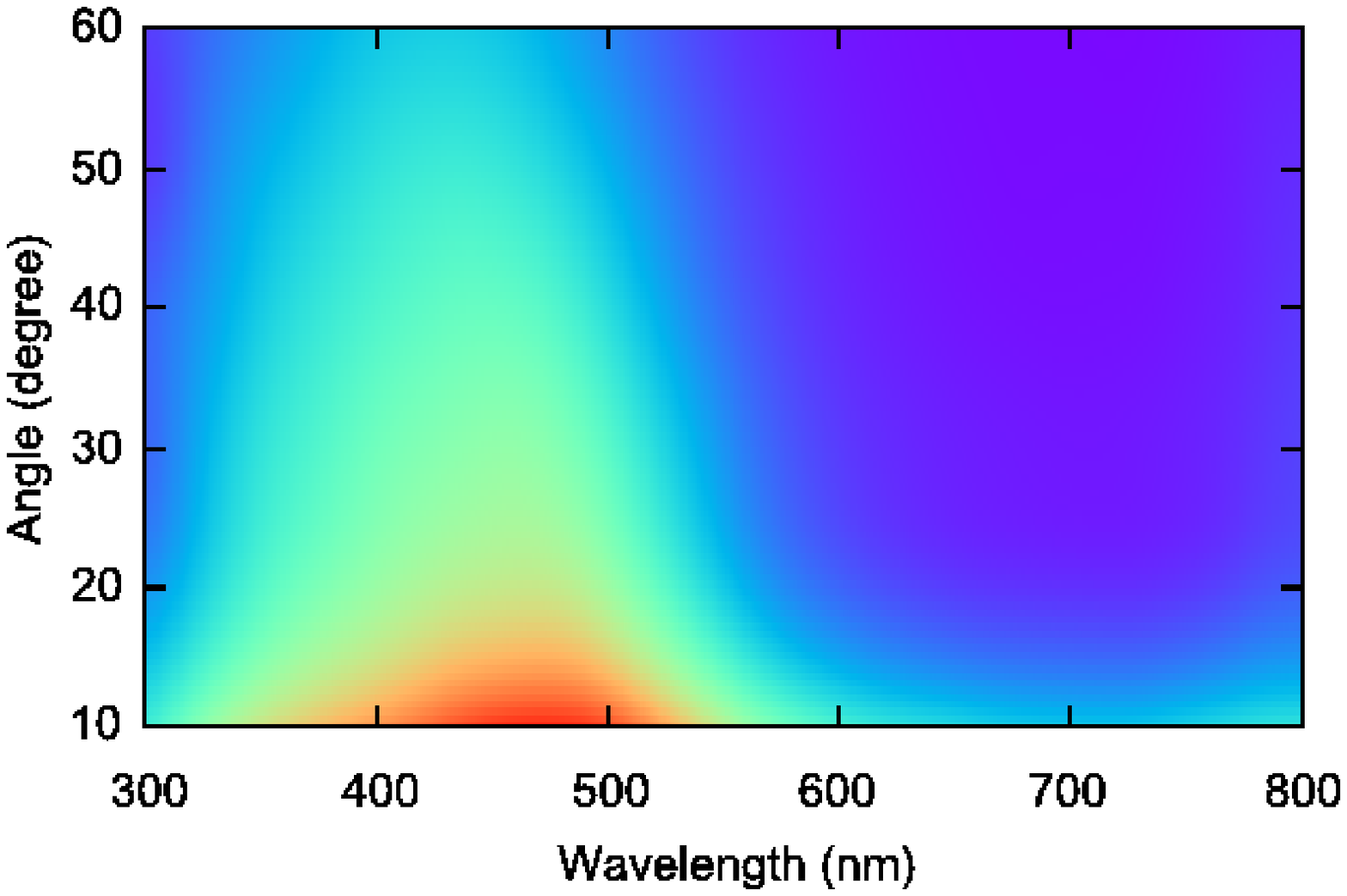}\\
(b)
\end{tabular}
\caption[Diffusive scattering]{False color maps showing the scattered light intensity as function of the observation angle $\theta$ and wavelength $\lambda$ from the feather barbs of {\it C. cotinga} (a), and {\it I. puella} (b). The experimental setup is shown in Fig. \ref{fig:setup} (c). The incident white light is normal to the barb surface, $\phi = 0$.}
\label{fig:diffusive}
\end{figure}

Finally we fix the sample while moving the detector [Fig. \ref{fig:setup}(c)]. The incident angle $\phi$ is kept at 0$^{\circ}$ (normal incidence), and the angle $\theta$ between the detection arm and the incident beam varies from 10$^{\circ}$ to 60$^{\circ}$. Figure \ref{fig:diffusive} shows the measured intensity as function of $\theta$ and $\lambda$. 
The spectral peaks can be observed at all $\theta$, confirming light is scattered into all directions. However there is a selectivity in the wavelength of scattered light, i.e., not all wavelengths are scattered with equal strength, leading to a peak in the scattering spectrum. The wavelength of scattering peak changes with $\theta$, suggesting the color depends on the observation angle. 

\begin{figure}[htbp]
\center
\includegraphics[width=3in]{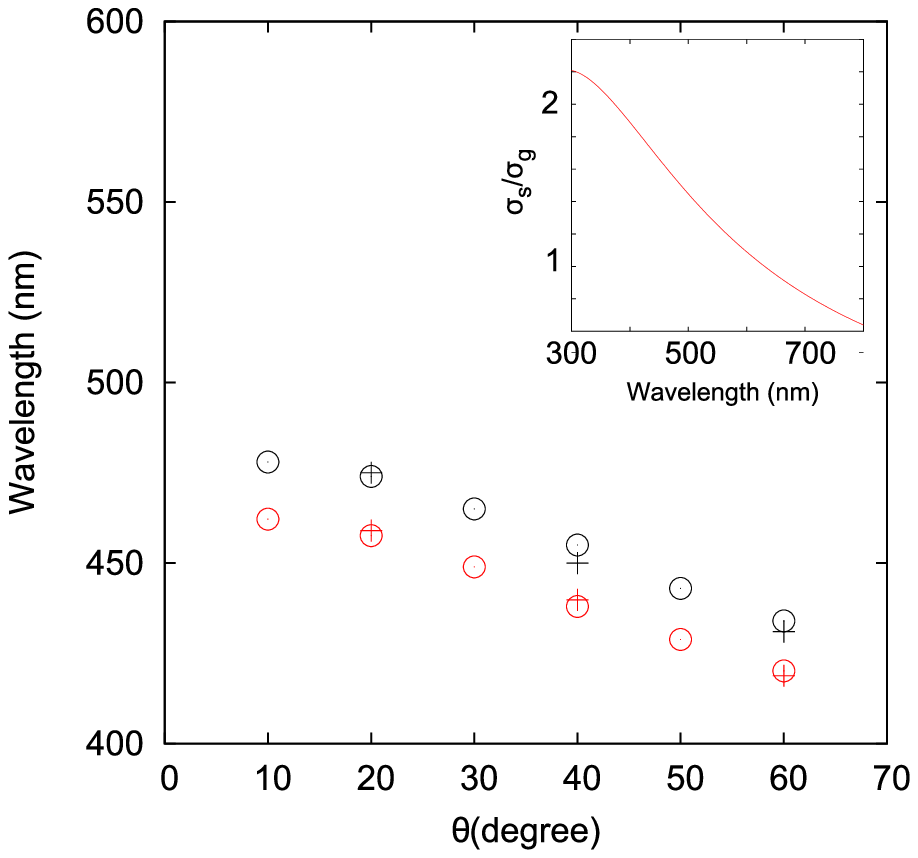}
\caption[wavelength vs. angle]{Wavelengths of specular reflection peaks (crosses) and scattering peaks (open circles) as a function of angle $\theta$ between the directions of illumination and observation. Red symbols for {\it C. cotinga}, and black for {\it I. puella}. Inset: calculated scattering cross section $\sigma_s$, normalized by the geometric cross section $\sigma_g$, of an air sphere embedded in $\beta$-keratin. The sphere diameter is 180 nm, close to the average diameter of air spheres in {\it C. cotinga}.}
\label{fig:q}
\end{figure}

A careful comparison reveals that the wavelength of light scattering peak at $\theta$ is equal to that of specularly reflected peak at $\phi = \theta/ 2$. In the specular reflection experiment [Fig. \ref{fig:setup}(a)], the angle between the incident light and the reflected light is $\theta = 2 \phi$. For all three experimental setups shown in Fig. \ref{fig:setup}, $\theta$ is the angle between the directions of illumination and observation.  In Fig. \ref{fig:q}, the wavelengths of reflection peaks and scattering peaks are plotted against $\theta$ for both types of quasi-ordered structures. It is evident that the peak wavelength depends only on $\theta$. Thus under directional lighting, the structural color depends not separately on the incident angle or the viewing angle, but on their difference that is the angle between the directions of illumination and observation. The same angular dispersion of reflection peak and scattering peak indicates reflection and scattering in the quasi-ordered structures have a common physical origin. In the following we include specular reflection in light scattering. 

\break

Since peak wavelength varies significantly with the observation angle, the scattering peak is not caused by Mie resonance of single particle scattering. For further confirmation, we calculate the scattering cross section $\sigma_s$ of a single air sphere embedded in $\beta$-keratin. The sphere diameter is $d$ =  180 nm, close to the average size of air spheres in the feather barbs of {\it C. cotinga}. The inset of Fig. \ref{fig:q} shows $\sigma_s / \sigma_g$  vs.  wavelength $\lambda$, where $\sigma_g$ is the geometric cross section of sphere ($\sigma_g = \pi d^2 / 4$). The refractive index of $\beta$-keratin is 1.58 \cite{brink_structural_2004}. In the wavelength range where a scattering peak is observed for {\it C. cotinga}, $\sigma_s / \sigma_g$ decays monotonically with increasing wavelength. This result confirms the scattering peak originates not from differential scattering of light wavelength by single particles, but from constructive interference of light scattered by many particles\cite{prum_coherent_1998}.

\begin{figure}[htbp]
\center
\includegraphics[width=3in]{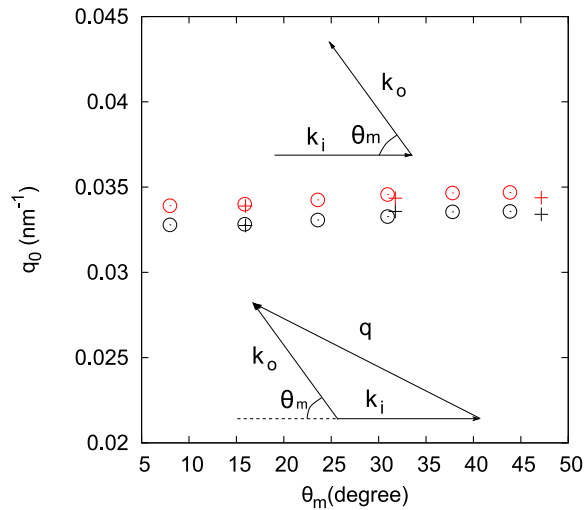}
\caption{Magnitude $q_0$ of spatial vectors corresponding to wavelengths of specular reflection peaks (crosses) and optical scattering peaks (open circles) as a function of angle $\theta_m$ between the directions of incident light and scattered light inside the sample. Red symbols for {\it C. cotinga}, and black for {\it I. puella}. Insets are single scattering diagrams. }
\label{fig:qm}
\end{figure}

Let us consider light with an incident wavevector ${\bf k}_i$ is scattered to a wavevector ${\bf k}_o$ (upper inset of Fig. Fig. \ref{fig:qm}). For elastic scattering, the magnitude of wavevector does not change.  $|{\bf k}_i| = |{\bf k}_o| \equiv k = 2 \pi n_e / \lambda$, where $n_e$ is the effective index of refraction, $\lambda$ is the vacuum wavelength.  The change in direction of wavevector is described by $\theta_m$, as shown in the lower inset of Fig. \ref{fig:qm}. This description is consistent with our experimental geometry. The difference between ${\bf k}_i$ and ${\bf k}_o$ is provided by the spatial variation of refractive index in the quasi-ordered nanostructure, ${\bf k}_o - {\bf k}_i = {\bf q}$, where ${\bf q}$ is the spatial vector of the structure. From the geometry in the lower inset of Fig. \ref{fig:qm}, we get 
\begin{equation}
2 k \cos(\theta_m / 2) = q.
\label{eq:single}
\end{equation} 

Fourier power spectrum of the refractive index distribution in space gives the spatial vectors existing in a structure. It is proportional to the SAXS pattern $I({\bf q})$. Since the quasi-ordered structure is isotropic, $I({\bf q})$ depend not on the direction of ${\bf q}$, but only on the magnitude $q \equiv |{\bf q}|$. As shown in Fig. \ref{fig:SAXS}, $I(q)$ is peaked at $q_0$, suggesting the dominant $q$ of a quasi-ordered structure is $q_0$. This allows us to substitute $q$ in Eq. (\ref{eq:single}) by $q_0$ and get
\begin{equation}
q_0 = {{4 \pi}\over{\lambda_0}} n_e \cos(\theta_m / 2).
\label{eq:single_1}
\end{equation}   
where $\lambda_0$ represents the wavelength at which scattering is the strongest, thus is equal to the peak wavelength of scattering spectrum. 

Eq. (\ref{eq:single_1}) shows that $\lambda_0$ must change with $\theta_m$ in order to keep $q_0$ constant. To verify this, we calculate the values of $q_0$ corresponding to the measured peak wavelengths of light scattering spectra. Optical refraction at the sample surface is taken into account to obtain $\theta_m$ inside the sample from the angle $\theta$ between the directions of illumination and observation outside the sample\cite{Noh2009a}. Figure \ref{fig:qm} is a plot of $q_0$ obtained from the experimental data. $q_0$ is the same for all $\theta_m$, consistent with the isotropic ring pattern of SAXS. The value of $q_0$ obtained from light scattering data coincides with that from SAXS data if we make $n_e = 1.25$. This corresponds to a volume fraction of $57\%$ for air in the quasi-ordered structures, which matches the estimates from electron micrographs. The good agreement of the predictions to the experimental data illustrates that the main peak in the optical scattering spectrum originates from single scattering, i.e. light is scattered only once inside the nanostructure. Note that single scattering is conceptually different from single particle scattering. The incident photons are scattered by many particles, although each photon is scattered at most once. The photons scattered by all the particles interfere constructively at the frequency of main scattering peak.    

A remaining question is why the optical scattering peak has larger amplitude for smaller $\theta$ as seen in Fig. \ref{fig:diffusive}. The invariance of $I({\bf q})$ with the direction of ${\bf q}$ seems to suggest that the amplitude of scattering peak should remain the same in all directions.  To find an answer, we calculated the scattered light intensity using $I({\bf q})$ from the SAXS data. To eliminate the artificial signal from diffraction by the beam block (inset of Fig. \ref{fig:SAXS}), we fit the curve of $I(q)$ for smaller $q$ (dashed line in Fig. \ref{fig:SAXS}), and used the fitted result for light scattering calculation. The polarization of light is ignored in the calculation. For single scattering process, the probability of incident light with ${\bf k}_i$ being scattered to ${\bf k}_o$ is proportional to $I({\bf q}={\bf k}_o-{\bf k}_i)$. Thus the flux $I_{\theta_m}$ of scattered photons in a unit solid angle around $\theta_m$ can be written as
\begin{equation}
I_{\theta_m}(k) = I(q) 2 \cos^2 \left( {{\theta_m}\over{2}} \right) = I \left[ 2k \cos \left( {{\theta_m}\over{2}} \right) \right] 2 \cos^2 \left( {{\theta_m}\over{2}} \right). 
\label{eq:single_2}
\end{equation}
The factor $2 \cos^2 (\theta_m / 2)$ comes from the relation between $k$ and $q$ in Eq. (\ref{eq:single}). 

\begin{figure}[htbp]
\center
\begin{tabular}{c c}
\includegraphics[width=3in]{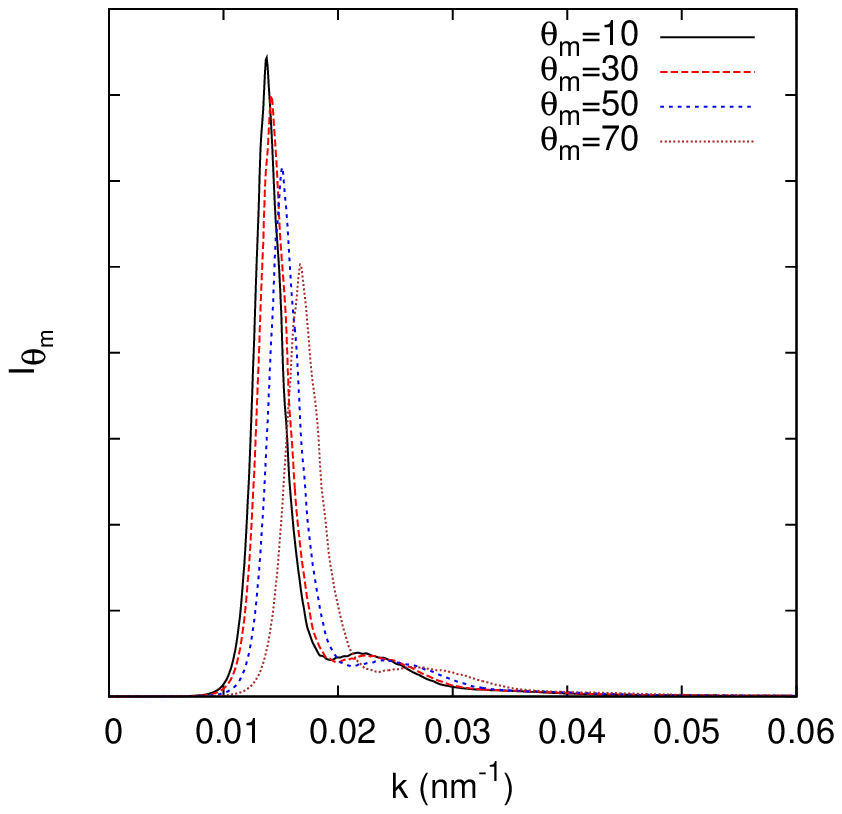}&
\includegraphics[width=3in]{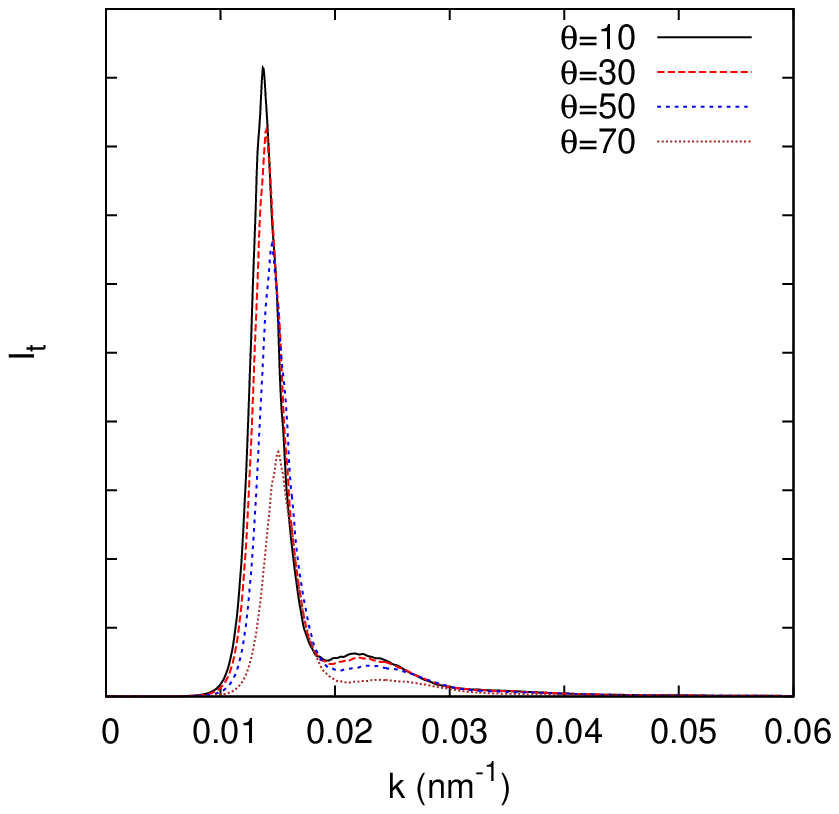}\\
(a)&(b)\\
\end{tabular}
\caption{Calculated light scattering spectra based on $I({\bf q})$ from the SAXS data of {\it C. cotinga}. (a) Calculated flux $I_{\theta_m}$ of light being scattered to a unit angle in the direction $\theta_m$ as a function of the magnitude $k$ of incident/scattered wavevectors. The values of $\theta_m$, the angle between the incident wavevector ${\bf k}_i$ and scattered wavevector ${\bf k}_o$, are shown in the legend. (b) Calculated flux $I_t$ of light being scattered to an angular range $\Delta \theta = 5^{\circ}$ around the direction $\theta$ as a function of $k$. The values of $\theta$, the angle between the directions of illumination and observation outside the sample, are shown in the legend.}
\label{fig:angle}
\end{figure}

Figure \ref{fig:angle}(a) plots $I_{\theta_m}$ as a function of $k$ for {\it C. cotinga}. 
As $\theta_m$ increases, the peak of $I_{\theta_m}(k)$ shifts to larger $k$. The ``speed'' of shift is not constant, at smaller $\theta_m$ the peak moves slowly with increasing $\theta_m$, then accelerates at larger $\theta_m$. This behavior can be explained by Eq. (\ref{eq:single_1}). For a constant $q$, e.g. $q= q_0$ at the peak,
\begin{equation}
\frac{\displaystyle dk}{\displaystyle d\theta_m}={{k}\over{2}} \tan \left( {{\theta_m} \over {2}} \right)
\end{equation} 
The change of $k$ with $\theta_m$ is proportional to $\tan (\theta_m/2)$. Near the backscattering direction, $\theta_m$ is close to 0, so is $\tan (\theta_m/2)$. $d k / d \theta_m \simeq 0$ means $k$ barely changes with $\theta_m$. Another feature in Fig. \ref{fig:angle}(a) is the reduction of peak height  with increasing $\theta_m$. This can be understood from the relation between $I_{\theta_m}(k)$ and $I(q)$ in Eq. \ref{eq:single_2}. Even though the peak of $I(q)$ has the same height for all angles, the value of $2 \cos^2 (\theta_m / 2)$ decreases with increasing $\theta_m$, thus lowering the peak height of $I_{\theta_m}(k)$ at larger $\theta_m$. 

Experimentally, the collection angle of scattered light is $\sim 5^\circ$. Thus we integrate $I_{\theta_m}(k)$ over the collection angle, corresponding to the addition of the curves in Fig. \ref{fig:angle}(a) within a range of $\theta_m$. Since the peaks shift more slowly with $\theta_m$ at smaller $\theta_m$, summation of the peaks over $\theta_m$ within the range of collection angle produces a bigger peak at smaller $\theta_m$ than that at larger $\theta_m$. This is confirmed in Fig. \ref{fig:angle}(b), which shows the integrated flux $I_t(k)$ of light being scattered to different directions described by $\theta$. The incident light (${\bf k}_i$) is normal to the barb surface, which is similar to the experiment in Fig. \ref{fig:setup}(c). As mentioned earlier, $\theta$ is different from $\theta_m$. For a constant collection angle in air, e.g. $\Delta \theta = 5^{\circ}$, the range of $\theta_m$ varies, namely, $\Delta \theta_m$ decreases with increasing $\theta_m$. It further lowers the peak of  $I_t(k)$ at larger $\theta$. Therefore, the peak height reduction in Fig. \ref{fig:angle}(b) is more dramatic than that in Fig. \ref{fig:angle}(a).

\section{non-iridescent Color under Omni-directional Illumination}

\begin{figure}[htbp]
\center
\begin{tabular}{c c}
\includegraphics[width=2in]{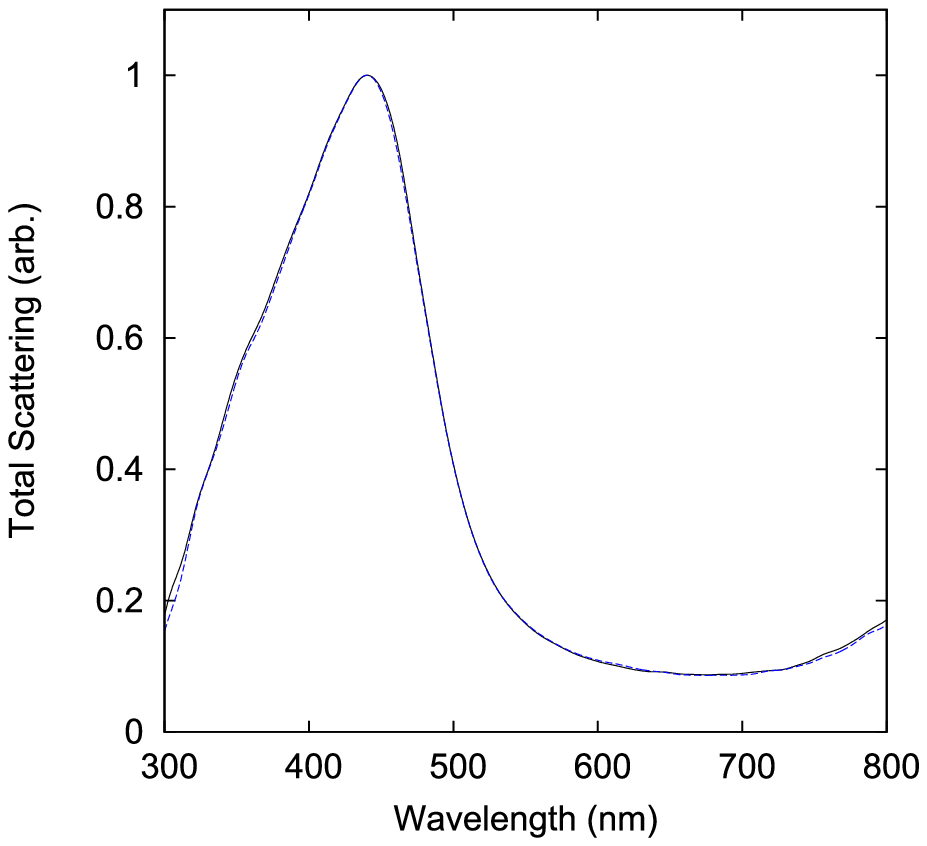}&
\includegraphics[width=2in]{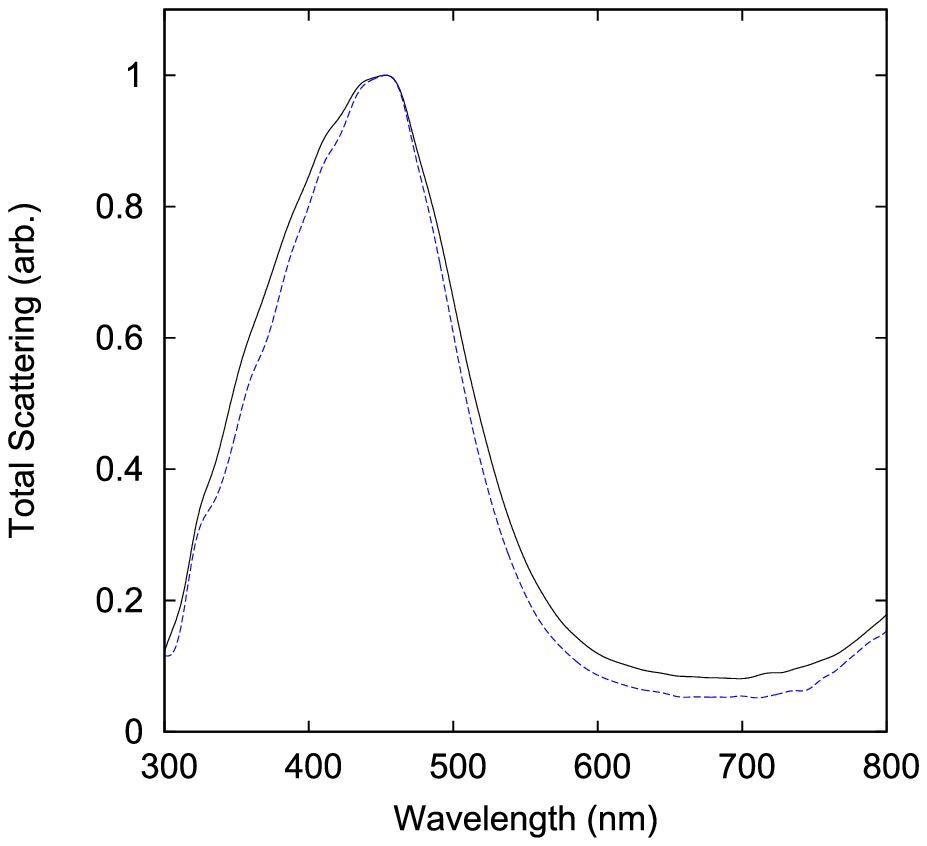} \\
(a)&(b)\\
\end{tabular}
\caption{Measured spectra of scattered light with omni-directional illumination of white light on feather barbs of {\it C. cotinga} (a) and {\it I. puella} (b). The angle between the normal of barb surface and the direction of observation $\psi = 0^{\circ}$ (black solid line), and $30^{\circ}$ (blue dashed line).}
\label{fig:total}
\end{figure}

\begin{figure}[htbp]
\center
\begin{tabular}{c c}
\includegraphics[width=2.5in]{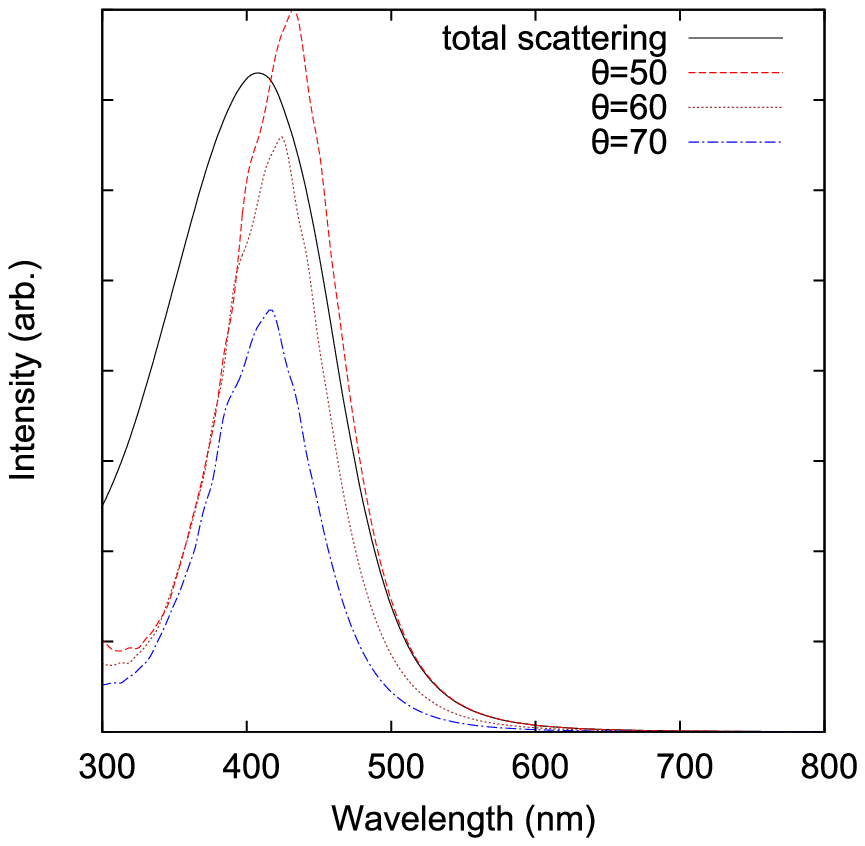}&
\includegraphics[width=2.5in]{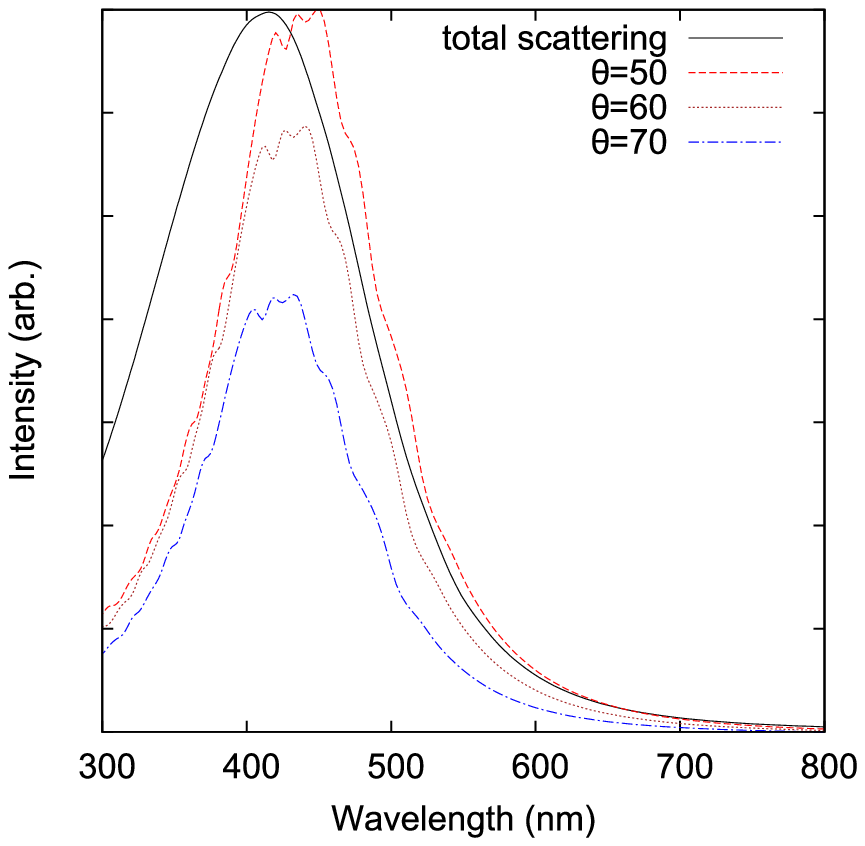}\\
(a)&(b)
\end{tabular}
\caption{Calculated spectra of scattered light under omni-directional illumination (black solid line) and directional illumination with $\theta = 50^{\circ}$ (red dashed line), $60^{\circ}$ (blue dotted line), and $70^{\circ}$ (green dash-dotted line) for {\it C. cotinga} (a) and {\it I. puella} (b). }
\label{fig:total_sim}
\end{figure}

Using the setup in Fig. \ref{fig:totalsetup}, we measured the spectra of scattered light at different viewing angles $\psi$ when the feather barbs were illuminated by white light in all directions. Figure \ref{fig:total} (a) and (b) show the spectra taken at  $\psi = 0^{\circ}, 30^{\circ}$ for the feather barbs of ${\it C. cotinga}$ and ${\it I. puella}$. For the quasi-ordered structures, the scattered light intensity varies drastically with wavelength. It demonstrates short-range order is sufficient to produce distinct color via constructive interference. Figures \ref{fig:total}(a) and (b) also show that the positions of scattering peaks do not change with $\psi$. The colors are invariant with the observation direction and thus non-iridescent. This phenomenon can be understood as that the structure is isotropic and when illumination is also isotropic all directions of observation are equivalent.  

To understand what determines the colors of quasi-ordered structures in the natural lighting condition, we calculate the spectra of scattered light under omni-directional illumination.  
Figure \ref{fig:total_sim} (a) and (b) show the calculated spectra, which are in good agreement with the data in Fig. \ref{fig:total} (a) and (b). For comparison,  we also plot in Fig.  \ref{fig:total_sim} the scattering spectra under directional illumination for several values of $\theta$. The scattering peak under omni-directional illumination is broader than that under directional illumination, because the former is a sum of the latter over many illumination angles. The peak wavelength under omni-directional illumination coincides with that under directional illumination with $\theta \sim 60^{\circ} - 70^{\circ}$. This means the color under omni-directional illumination is not determined by the wavelength of backscattering peak, despite the scattering intensity being the strongest in the backward direction. This is because the solid angle $\Omega$ corresponding to a fixed range of $\theta$  is proportional to $\sin\theta$. For larger $\theta$, the solid angle available for lighting is larger, producing more scattered light at shorter wavelength. A balance between the decrease of scattered light intensity and increase of available solid angle with $\theta$  makes the scattering spectra under omni-directional lighting peaked at the wavelength of that under directional lighting with $\theta \sim 60^{\circ} - 70^{\circ}$.

\section{Conclusion}

We investigated the mechanism of color production by quasi-ordered nanostructures in bird feather barbs. There are two types of such structures: the channel type and the sphere type. Small-angle X-ray scattering (SAXS) patterns from both exhibit rings, implying the nanostructures are isotropic and have strong correlations at particular length scales. {\it In-situ} SAXS data confirmed the structures are ordered on the length scale comparable to optical wavelength, and the order is short-ranged. 

We conducted angle-resolved reflection and scattering spectrometry to fully characterize the colors produced by the two types of nanostructures with short-range order. Under directional lighting, the spectra of reflected and scattered light vary with the incident angle and viewing angle. Our data demonstrated that the reflection/scattering peak wavelengths depend only on the angle between the directions of illumination and observation. Such dependence can be well explained by single scattering of light in an isotropic structure with short-range order. From the dominant length scale of structural correlation, we predicted the angular dispersions of reflection/scattering peaks which agree well with the experimental data. Using the Fourier power spectra of structure from the SAXS data, we calculated optical scattering spectra at various angles and explained why the scattering peak is the highest in the backscattering direction (opposite to the incident light direction). 

To study coloration in the natural lighting condition, we illuminated the feather barbs with white light from all directions and compared the coloration to that in the directional lighting condition. Under omni-directional illumination, colors from the quasi-ordered structures are invariant with the viewing angle and thus non-iridescent. The scattering peak is broader than that with directional lighting. Its center wavelength is shorter than that of backscattering peak, indicating the color under natural light is not determined by the backscattering. Despite the backscattering intensity is the strongest, the available solid angle for lighting is the smallest in the backscattering direction. We calculated single  scattering spectra under omni-directional illumination and obtained good agreement with the measurement results. Our studies demonstrated the main peaks in light scattering/reflection spectra originate from constructive interference of light that is scattered only once by the quasi-ordered nanostructures in the bird feather barbs.



In summary, the quasi-ordered structures in the bird feather barbs utilize the short-range order to produce distinct colors via constructive interference of scattered light. The isotropic nature of quasi-ordered structures makes the colors non-iridescent under natural light.  

 
We thank Drs. Yongle Pan and Yidong Chong for many useful discussions.
This work was supported with seed funding from the Yale NSFMRSEC (DMR-0520495) and NSF grants to HC (ECCS-0823345), ERD (CBET), SGJM (DMR), and ROP (DBI). Feather specimens were provided by the Yale Peabody Museum of Natural History and the University of Kansas Natural History Museum and Biodiversity Research Center. TEMs of feathers were prepared by Tim Quinn. SAXS data were collected at beam line 8-ID-I at the Advanced
Photon Source at Argonne National Labs with the help of Drs. Alec Sandy and Suresh Narayanan, and supported by the U. S. Department of Energy, Office of Science, Office of Basic Energy Sciences, under Contract No. DE-AC02-06CH11357.

\bibliographystyle{unsrt}
\bibliography{birdSC}

\end{document}